\documentclass[aps,prl,reprint,amsmath,amssymb,longbibliography]{revtex4-2}
\usepackage{amsmath}
\usepackage{amssymb}
\usepackage{physics}
\usepackage{graphicx}
\graphicspath{ {./images/} }
\usepackage[
    colorlinks=true,
    linkcolor=blue,
    citecolor=blue,
    urlcolor=blue
]{hyperref}

\usepackage{mathtools}

\begin{document}

\title{Black hole radial geometry from a quantum spin chain}

\author{Boris A. Khanikati}
\email{bkhanikaev@g.harvard.edu}
\affiliation{%
 Department of Physics, Harvard University, Cambridge MA 02138, USA
}%

\date{\today}

\begin{abstract}
We show that the radial differential equation for a scalar field in an anti-de Sitter black brane background can be encoded in an exactly solvable semi-infinite spin chain independent of dimensionality. The radial coordinate emerges as lattice quasimomentum. Spin coherence encodes the geometry, while the spectrum and response recover gravitational observables and yield exact screening-mass sum rules. We implement the finite chain on a superconducting processor. At finite filling and low energies, the model yields a conformal field theory in a coordinate proportional to proper radial distance.
\end{abstract}

\maketitle


\textit{Introduction.---}A key lesson of gauge-gravity duality is that spacetime geometry can be encoded in quantum degrees of freedom, often through intrinsically many-body structures such as entanglement \cite{maldacena1999adsCft, gubser1998gauge, witten1998ads, ryu2006holographic, hubeny2007covariant, vanRaamsdonk2010, maldacena2013coolHorizons}. This has prompted the search for simple quantum systems with emergent geometry. The Sachdev-Ye-Kitaev (SYK) model provides a prominent example through its low-energy relation to Jackiw-Teitelboim (JT) gravity in $\rm AdS_2$ \cite{sachdev1993syk, kitaev2015syk, sachdev2015syk, maldacena2016syk, xu2020syk, jafferis2022wormhole}, however its all-to-all interactions remain challenging to realize. On the other hand, quantum devices can engineer controlled spin Hamiltonians and quantum field theories \cite{georgescu2014simulation, Browaeys2020, monroe2021simulation}. This motivates representing spacetime geometry in simple spin systems amenable to experimental implementation \cite{kristjansen2026blackhole, daniel2025scrambling, Basteiro2022}.

In this Letter, we show that the radial scalar equation in an $\mathrm{AdS}_{d+1}$ black brane background admits an exact representation in the one-magnon sector of a semi-infinite nearest-neighbor spin chain. The chain is exactly solvable and its Hamiltonian is independent of dimensionality. The exterior radial coordinate emerges as an asymptotic lattice quasimomentum and can be reconstructed from coarse-grained spin correlations, while analytic continuation extends the dictionary into the black hole interior. The mapping reproduces the Breitenlohner-Freedman (BF) bound and thermal one-point coefficient, encodes the interior proper time, and yields the massless-scalar screening spectra and exact screening-mass sum rules. We implement the finite chain on a superconducting quantum processor and reconstruct the proper time to the singularity with an $\sim8\%$ deviation from the exact value. Finally, upon second quantization, the spectral edge becomes a many-body quantum critical point, while at finite filling the low-energy theory is a $c=1$ conformal field theory (CFT) whose natural spatial coordinate is proportional to the black brane proper radial distance, although the gravitational interpretation of this CFT remains to be understood.

\begin{figure*}[t!]
    \centering
    \includegraphics[trim=0 1cm 0 0cm, width=\linewidth]{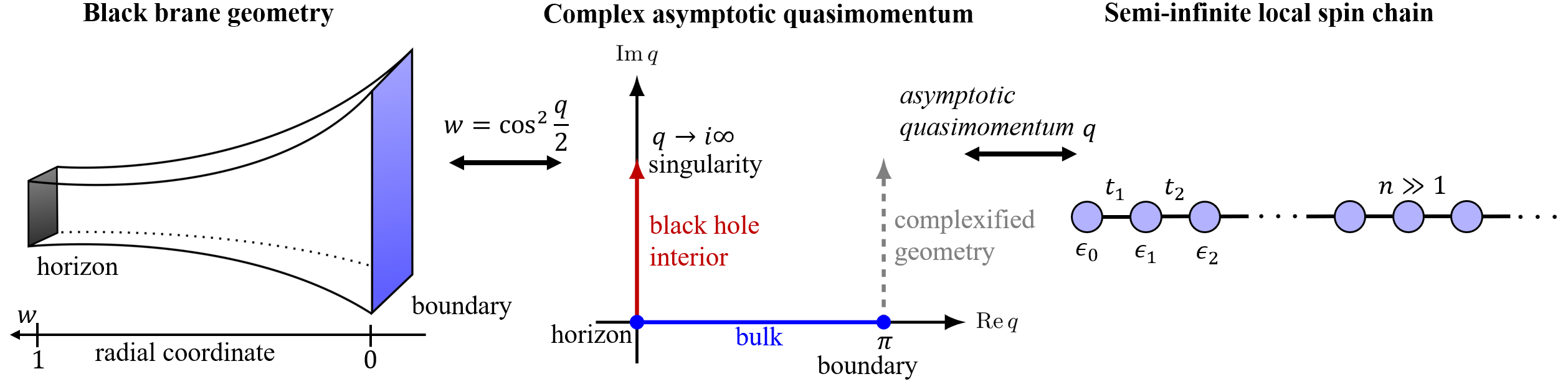}
    \caption{Schematic representation of the dictionary between the black brane geometry and the spin chain. The dimensionless radial coordinate $w$ runs from the $d$-dimensional boundary ($w=0$) to the horizon ($w=1$), with $w>1$ corresponding to the black hole interior. In the asymptotic regime, the radial coordinate is mapped onto the spin-chain quasimomentum. Real $q\in(0,\pi)$ describes the black brane bulk, whereas its analytic continuation, $q=i\kappa$ with $\kappa>0$, describes the black hole interior.}
    \label{fig:fig_1}
\end{figure*}

\textit{Black brane as a spin chain.---}We consider an AdS black brane, which features a planar event horizon. The $\mathrm{AdS}_{d+1}$ metric is given by
\begin{equation}
    ds^2=\frac{R^2}{z^2}\left[-f(z)dt^2+\frac{dz^2}{f(z)}+d\mathbf{x}_{d-1}^2\right],
\end{equation}
where $f(z)=1-(z/z_0)^d$, $R$ is the $\rm AdS$ radius, $z=0$ is the boundary, and $z=z_0$ is the event horizon, whose Hawking temperature is $T=d/(4\pi z_0)$. Consider a scalar field $\phi$ of mass $m$, dual to a boundary operator of scaling dimension $\Delta$, with the usual relation $m^2R^2 = \Delta(\Delta - d)$. It is useful to introduce the dimensionless variables $w=\left(z/z_0\right)^d, \quad h = \Delta/d$. Then, $w \in (0, 1)$ parameterizes the AdS geometry, with $w = 0$ as the boundary and $w = 1$ as the horizon. For the static homogeneous mode, the sourced Klein--Gordon equation reduces to \cite{berenstein2023ads}
\begin{align}
    [L + h(h-1)]\phi(w) = - B_dw^2, \nonumber \\    L = -w^2 \frac{d}{dw}\left[(1- w)\frac{d}{dw} \right]. 
\end{align}
The coefficient $B_d$ contains the strength and normalization of the curvature coupling. Importantly, the operator $L$ contains no explicit dependence on the dimensionality.

With the inner product $(f,g)=\int_0^1dw\,f^*(w)g(w)/w^2$, $L$ is self-adjoint, so let us choose to associate with it the Hamiltonian of a system. We wish to construct a basis in which this Hamiltonian will be nearest-neighbor. Since the source term is $w^2$, applying Gram-Schmidt to $w^2,w^3,\ldots$ gives the orthonormal basis $\psi_n=\sqrt{2n+3}\,w^2P_n^{(0,2)}(2w-1)$. In this basis, $L$ is tridiagonal. Identifying $\ket n$ with $\psi_n$ gives the on-site terms and nearest-neighbor coupling amplitudes, 
\begin{equation}
    \epsilon_n = \frac{(n+1)(n+2)}{2}, \quad  t_{n+1} = \frac{(n+1)(n+2)(n+3)}{2\sqrt{(2n+3)(2n+5)}}. 
\end{equation}

As such, the problem may be represented as a semi-infinite spin chain $n=0,1,2,\ldots$. With the fully polarized state $\ket\Omega=\ket{\downarrow\downarrow\cdots}$ as the vacuum and $\ket n=\sigma_n^+\ket\Omega$, the Hamiltonian is
\begin{equation}\label{eq:main-hamiltonian}
    H = \sum_{n = 0}^\infty \epsilon_n \sigma_n^+ \sigma_n^- + \sum_{n = 0}^\infty t_{n+1} (\sigma_{n+1}^+ \sigma_n^- + \sigma_{n}^+ \sigma_{n+1}^{-}).
\end{equation}
Thus, the radial scalar problem is encoded in the one-magnon sector of a nearest-neighbor, number-conserving spin chain with inhomogeneous couplings. The resulting Hamiltonian is the same for every planar $\mathrm{AdS}_{d+1}$ Schwarzschild geometry, with the dimensional dependence entering through the relation of $w$ and $h$ with $d$.

\textit{Radial geometry and quasimomenta.---}We next ask how the radial coordinate $w$ is encoded in the lattice representation. Fixing $w$ gives the amplitude $\{\psi_n(w)\}$ across the lattice. Parameterize $0<w<1$ by $q\in(0,\pi)$ through
\begin{equation}\label{eq:dictionary}
    w=\cos^2 \frac{q}{2}.    
\end{equation}
Using the large-degree asymptotics of Jacobi polynomials, we find that for large $n$,
\begin{equation}\label{eq:large-n}
    \psi_n(w) \simeq \sqrt{\frac{2}{\pi}} \frac{\cos^{3/2}(q/2)}{\sin^{1/2}(q/2) }\cos\left[\left(n + \frac{3}{2} \right) q - \frac{\pi}{4} \right].
\end{equation}
A radial position $w$ is therefore represented asymptotically by an oscillation in $n$ with quasimomentum $q(w)$, Eq.~(\ref{eq:dictionary}). The horizon corresponds to $w\to1$, $q\to0$, while the boundary corresponds to $w\to0$, $q\to\pi$. A schematic representation of this dictionary is shown in Fig.~\ref{fig:fig_1}.

It is convenient to introduce a dimensionless exterior proper distance $\rho$ by $w=\sech^2(\rho/2)$, $\ell=R\rho/d$, where $\ell$ is the proper radial distance measured outward from the horizon along a constant-$t$, constant-$\mathbf x$ slice. 

\textit{Radial geometry from spin correlations.---}The radial coordinate can also be extracted from the spin coherence $C_{ij}=\langle\sigma_i^+\sigma_j^-\rangle$. Since $\{\psi_n(w)\}_{n \geq 0}$ is not normalizable, we define the normalization-independent, coarse-grained ratio
\begin{equation}
    \mathcal G_r(w) = \lim_{\substack{n_0\to\infty\\L\to\infty}} \frac{\sum_{n=n_0}^{n_0+L-1}\psi_n(w)\psi_{n+r}(w)} {\sum_{n=n_0}^{n_0+L-1}\psi_n^2(w)}.
\end{equation}
Using Eq.~(\ref{eq:large-n}) gives $\mathcal G_r=\cos(rq)$. In particular,
\begin{equation}
    w=\frac{1+\mathcal G_1}{2}, \quad \ell(\mathcal G_1)=\frac{2R}{d}\sinh^{-1} \sqrt{\frac{1-\mathcal G_1}{1+\mathcal G_1}}.
\end{equation}

Thus, a measurement of the coarse-grained nearest-neighbor coherence fixes the corresponding radial coordinate. Given the external parameters $d$, $R$, and $z_0$, the same observable can therefore be used to reconstruct the radial metric profile,
\begin{align}
    ds^2 & = \frac{R^2}{z_0^2} \left( \frac{2}{1+\mathcal G_1}\right)^{2/d} \left[ -\frac{1-\mathcal G_1}{2}\,dt^2 +d\mathbf{x}_{d-1}^2 \right] \nonumber\\
    &\quad +  \frac{2R^2}{d^2} \frac{(\mathrm d\mathcal G_1)^2}{(1-\mathcal G_1)(1+\mathcal G_1)^2}.
\end{align}
The endpoints $\mathcal G_1=1$ and $\mathcal G_1=-1$ correspond to the horizon and the AdS boundary, respectively; varying $\mathcal G_1$ reconstructs the radial metric profile.

\textit{Black hole interior.---}To continue through the horizon, we extend the radial coordinate to $w>1$ (see Fig.~\ref{fig:fig_1}) by analytically continuing the quasimomentum to $q=i\kappa$, with $\kappa\geq0$. The radial dictionary then becomes $w=\cosh^2(\kappa/2)$, with $\kappa=0$ corresponding to the horizon and $\kappa\to\infty$ to the singularity. Requiring $w$ to remain real also admits the branch $q=\pi+i\kappa$, for which $w\leq0$. This branch lies outside the original real Lorentzian black brane geometry, and we leave its possible significance for future investigation.

Inside the horizon, the radial direction becomes timelike. At fixed $t$ and $\mathbf{x}$, $d\tau_{\rm in}=(R/d)\,dw/[w\sqrt{w-1}]$. Writing $w=\sec^2(\theta/2)$ gives $\tau_{\rm in}=R\theta/d$, where $\theta=0$ at the horizon and $\theta\to\pi$ at the singularity. Hence, the proper time from the horizon to the singularity is
\begin{equation}\label{eq:prop-time}
    \tau_s= \frac{\pi R}{d}.
\end{equation}
This reproduces the result obtained from thermal one-point functions~\cite{grinberg2021properTime}.

\textit{Breitenlohner-Freedman bound.---}Let us now solve the associated problem. Expanding $\ket E=\sum_{n=0}^{\infty}u_n(E)\ket n$ gives
$t_nu_{n-1}+\epsilon_nu_n+t_{n+1}u_{n+1}=Eu_n$, which is the recurrence relation of Wilson polynomials with parameters $(1/2,1/2,3/2,3/2)$. Consequently, the spectrum is 
$\mathrm{spec}(H)=[1/4,\infty)$. Explicit eigenfunctions and the spectral measure are given in Appendix A.

For the homogeneous scalar equation, a solution corresponds to energy $E=-h(h-1)$, where $h=\Delta/d$. For real $h$, $E(h) \leq 1/4$ and reaches the lower continuum edge of the chain at $h = 1/2$, or $\Delta = d/2$. Together with the mass-dimension relation $m^2R^2=\Delta(\Delta-d)$, the threshold gives
\begin{equation}
    m^2R^2=-\frac{d^2}{4},
\end{equation}
which is the Breitenlohner-Freedman (BF) stability bound in $\mathrm{AdS}_{d+1}$ \cite{breitenlohner1982a,breitenlohner1982b}. Thus, the spectral edge of the chain represents the BF stability bound. 

\textit{Thermal one-point function.---}For a scalar dual to an operator $\mathcal O$ of dimension $\Delta = dh$, the near-boundary solution behaves as $\phi(w) \sim Jw^{1 - h} + cw^h$. For the thermal one-point function considered here, $J = 0$ and the Weyl tensor supplies the source $B_d w^2$. Since $\psi_0(w) = \sqrt{3}\,w^2$ by construction, this source is localized at the first lattice site. 

The spectral decomposition gives the resolvent $G(\zeta)=(H-\zeta)^{-1}$, whose matrix elements are
\begin{equation}\label{eq:Gmn}
    G_{mn}(\zeta)=\int_0^\infty d\lambda\,
    \frac{\varrho(\lambda)p_m(\lambda^2)p_n(\lambda^2)}
    {\frac14+\lambda^2-\zeta},
\end{equation}
where $\varrho(\lambda)$ is the normalized spectral density given explicitly in Appendix A of the End Matter.
Its branch cut begins at the spectral edge $\zeta=1/4$, which, as shown above, corresponds to the BF bound. Writing $s=h-1/2$, the scalar field solution is
\begin{equation}\label{eq:ket-phi}
    \ket{\phi}=- \frac{B_d}{\sqrt 3} G\left( \frac{1}{4} - s^2 \right) \ket 0.
\end{equation}
For $s>0$, the resolvent is evaluated below the continuum edge. Because temperature provides the only scale, the dependence is fixed as $\langle\mathcal O\rangle\propto z_0^{-\Delta}\propto T^\Delta$ \cite{grinberg2021properTime}, leaving only the coefficient to be determined. In the spin chain, this coefficient $c(h)$ is obtained from the alternating large-$n$ boundary component of Eq.~(\ref{eq:ket-phi}), since the AdS boundary corresponds to $q\to\pi$.

The projection of the boundary mode onto our basis is, for large $n$, 
\begin{equation}
    b_n(h) = (\psi_n, w^h) \simeq (-1)^n \sqrt{2} \frac{\Gamma(h+1)}{\Gamma(2 - h)}n^{1/2 - 2h}.  
\end{equation}
The asymptotics of the Wilson polynomials give
\begin{equation}
    G_{n0}\left(\frac{1}{4} - s^2 \right) \simeq (-1)^n \sqrt{6} \frac{\Gamma(h)^2 \Gamma(h+1)^2}{\Gamma(2h)}n^{1/2 - 2h}. 
\end{equation}
Matching these amplitudes therefore determines the normalizable coefficient,
\begin{equation}\label{eq:one-point}
    c(h) = -B_d\pi \frac{\Gamma(h)^2}{\Gamma(2h)}\frac{h(1-h)}{\sin(\pi h)},
\end{equation}
in agreement with the expected result \cite{berenstein2023ads}. Generically, the one-point function takes the form $\langle\mathcal O\rangle=\mathcal N_\Delta c(h)z_0^{-\Delta}$, where $\mathcal N_\Delta$ contains the overall normalization.

Following Ref.~\cite{grinberg2021properTime}, the phase of the large-mass thermal one-point function encodes the interior proper time. Defining
\begin{equation}
    \mathcal R(h)=c(h)\frac{\Gamma(2h)}{\Gamma(h)^2h(1-h)} = -\frac{\pi B_d}{\sin(\pi h)},
\end{equation}
and analytically continuing to $h=x-i\gamma$, the unwrapped phase has mean winding rate $-\pi$ per unit $x$. Since $x-1/2\simeq mR/d$ at large mass, this winding gives $\tau_s=\pi R/d$, in agreement with Eq.~(\ref{eq:prop-time}).

\textit{Screening spectrum and sum rules.---}At large separation, the connected correlator decays as $\langle\mathcal O(\mathbf{x})\mathcal O(0)\rangle_c\sim e^{-M|\mathbf{x}|}$, where $M$ is the inverse correlation length known as the screening mass. For a massless scalar $\Phi(z,\mathbf{x})=e^{-Mx_1}\phi(z)$ in the planar AdS$_{d+1}$ black brane, the radial equation becomes
$H\phi = (Mz_0/d)^2w^{2/d}\phi$. Let $W$ denote multiplication by $w$ on the radial Hilbert space. The radial equation is then the generalized eigenvalue problem
\begin{equation}\label{eq:screening-mass}
    4H\ket{\phi_j}=\Lambda_j W^{2/d}\ket{\phi_j}, \quad \Lambda_j=\left(\frac{M_j}{2\pi T}\right)^2 .
\end{equation}
Thus, the massless-scalar screening spectrum for arbitrary $d$ is determined by the same operators $H$ and $W$, with dimensional dependence entering only through $W^{2/d}$. After truncation, the problem reduces to finite-dimensional linear algebra, while inverse spectral moments become traces of a compact operator.

To the best of our knowledge, these sum rules for the screening spectrum have not been established previously. Reformulating the generalized eigenvalue problem as an ordinary eigenvalue problem introduces the operator $\mathcal K_d=(4H)^{-1/2}W^{2/d}(4H)^{-1/2}$, whose eigenvalues are $\Lambda_j^{-1}$. Its spectral traces therefore give the hierarchy
\begin{equation}
    \sum_{j=0}^{\infty}\Lambda_j^{-p} = \Tr\mathcal K_d^p = \sum_{j=0}^{\infty} \left(\frac{2\pi T}{M_j}\right)^{2p}.
\end{equation}
Using the inverse of $H$, the first inverse moment evaluates to
\begin{equation}\label{eq:sum-rule}
    \sum_{j=0}^{\infty} \left(\frac{2\pi T}{M_j}\right)^2 = \frac{\psi(2/d) + \gamma_{\mathrm E}}{4(2/d-1)},\quad d\neq2,
\end{equation}
where $\psi(x)$ is the digamma function and $\gamma_{\mathrm E}$ is the Euler-Mascheroni constant. For $d=4$, this reduces to the particularly simple exact relation
$\sum_{j=0}^{\infty}(2\pi T/M_j)^2=\log 2$. As checks, the same eigenproblem reproduces the known $d=4$ lowest massless-scalar screening mass~\cite{bak2007screening} and the exact nonrotating Ba\~nados--Teitelboim--Zanelli (BTZ) ($d=2$) spectrum $M_j/\pi T=4(j+1)$~\cite{birmingham2002cft}. It also reproduces the large-$j$ WKB spacing~\cite{Minahan1999}.

\textit{Many-body extension.---}The one-magnon construction extends naturally to a many-body system. Under a Jordan-Wigner transformation, the spin Hamiltonian becomes the inhomogeneous free-fermion chain
\begin{equation}
    H = \sum_{n\geq0}\epsilon_n c_n^\dagger c_n + \sum_{n\geq0}t_{n+1}  \left(c_{n+1}^\dagger c_n + c_n^\dagger c_{n+1}\right),
\end{equation}
whose one-particle sector is the Hamiltonian discussed above. Introducing a chemical potential, $H_\mu=H-\mu\hat N_f$, shifts the one-particle spectrum to $E_\mu(\lambda)=\frac{1}{4}+\lambda^2-\mu$. At fixed chemical potential, occupation begins at $\mu_c=1/4$. At this threshold the excitations have quadratic dispersion, $E_{\mu_c}\sim\lambda^2$, giving a $z=2$ quantum critical point. Thus, the same spectral edge that encodes the BF bound in the one-particle problem becomes a many-body quantum critical point.

To describe the finite-density low-energy theory, consider the thermodynamic limit at fixed filling, $N_f=\nu N$, with $u=n/N$ and a chemical potential scaling as $\mu_N=a^2N^2$, where $a$ is fixed by the filling. The large-$n$ form of the lattice Hamiltonian gives the semiclassical one-particle energy
\begin{equation}
\mathcal E(u,q)=u^2\cos^2\frac{q}{2} = u^2 w(q).
\end{equation}
At the Fermi surface, $\mathcal E=a^2$. Using the radial dictionary equation~(\ref{eq:dictionary}), the Fermi surface determines the radial profile
\begin{equation}
\rho_F(u)=2\operatorname{arcosh}\frac{u}{a}, \quad u \geq a.
\end{equation}
For $u<a$, the local band is completely filled, so the gapless region is $a<u<1$.

At fixed finite filling, the low-energy level spacings scale as $N$, so we take the continuum limit of $H_{\mu_N}/N$. Linearizing about the Fermi surface away from the turning point $u=a$ gives a $c=1$ free-fermion CFT. The continuum derivation, including the relation between $a$ and the filling, is given in Appendix B of the End Matter. The local Fermi velocity is $v_F(u)=a\sqrt{u^2-a^2}$ for $u\geq a$. Because the Fermi velocity varies along the chain, the natural coordinate of the low-energy CFT is
\begin{equation}
\xi(u)=\int^u\frac{du'}{v_F(u')} = \frac{1}{a}\operatorname{arcosh}\frac{u}{a} = \frac{\rho_F(u)}{2a}.
\end{equation}
In the coordinate $\xi$, the position dependence of the Fermi velocity is absorbed and the low-energy theory takes the locally homogeneous form of a $c=1$ free-fermion CFT. Since the proper radial distance of the black brane is $\ell=R\rho/d$, it follows that
\begin{equation}
\xi=\frac{d}{2aR}\ell.
\end{equation}
Thus, the natural spatial coordinate of the resulting $c=1$ CFT is proportional to the black brane proper radial distance.

\textit{Experimental implementation.---}The one-particle experiment establishes the hardware realizability of the construction and serves as a platform for probing interacting extensions, although the implementation itself does not provide a quantum computational advantage. We truncate the semi-infinite chain to $N=32$ sites and implement the resulting dynamics on the 156-qubit IBM \texttt{ibm\_marrakesh} superconducting quantum processor.

To probe the response to the curvature source, which is localized at the first lattice site, we prepare the state $\ket{\Psi(0)}=(\ket{\Omega}+\ket0)/\sqrt2$. Because the vacuum is invariant under the number-conserving evolution, it provides a phase reference for the one-magnon component. 
The propagator is therefore accessible through single-qubit observables, $U_{n0}(\tau) = \langle X_n(\tau)\rangle + i\langle Y_n(\tau)\rangle$. Thus, two global measurement settings at each evolution time determine the propagator $U_{n0}(\tau)$ across the chain. 

The measured real-time dynamics is used to reconstruct the static radial response and the thermal one-point coefficient $c(h)$. Under analytic continuation $h=x-i\gamma$, the phase of $\mathcal R(h)$ encodes the proper time from the horizon to the singularity. For $d = 4$, a linear fit to the experimental phase gives $\tau_s/R=0.847\pm0.017$, compared with the exact black brane value $\pi/4\simeq0.785$, corresponding to a $\sim7.8\%$ relative deviation. The quoted uncertainty reflects finite-shot statistics only and does not include systematic hardware, finite-size, finite-time, or reconstruction errors. Further details of the circuit implementation, reconstruction procedure, and experimental data are given in the Supplemental Material.

\textit{Discussion and outlook.---}A key feature of the construction is that it provides a setting in which the gravitational dictionary can be probed directly with experimentally accessible quantum degrees of freedom. The exact screening sum rules further show that the lattice representation is not restricted to reproducing known gravitational quantities, but can also reveal structure that is less apparent in the continuum formulation.

We have focused here on static observables. A natural next step is to include finite-frequency bulk modes and determine whether the resulting radial problem retains a comparably simple lattice representation. Such an extension could provide access to dynamical quantities, including transport coefficients and frequency-dependent correlation functions.

A more ambitious question is whether gravitational interactions can also be encoded in the spin system. The present construction describes scalar fields in a fixed black brane geometry; incorporating backreaction would require the encoded geometry, and hence the effective lattice description, to respond dynamically to the matter sector. This would require an interacting or coupled quantum system beyond the fixed radial chain considered here. Whether such a construction can reproduce even a restricted form of gravitational backreaction remains an open question, but would represent a step toward encoding gravitational dynamics itself.

A complementary direction concerns the many-body extension. At fixed finite filling, its low-energy limit in the gapless region is a $c=1$ CFT whose natural spatial coordinate is proportional to the proper radial distance of the black brane. It would be interesting to determine whether other many-body observables, including entanglement and dynamical correlations, admit similarly direct geometric interpretations. Independently of its gravitational interpretation, the resulting inhomogeneous free-fermion model provides an unusual realization of conformal physics, with couplings that grow quadratically with lattice position and, asymptotically, scale transformations of the lattice coordinate becoming translations in the emergent conformal coordinate. This structure may provide a useful setting for studying quantum criticality and nonequilibrium dynamics in inhomogeneous quantum systems.

\begin{acknowledgments}
I thank Enes Banushi, Ayush Gilotra, Alexander Khanikaev, Kason Kunkelmann, Subir Sachdev, Sanjit Shirol, Zheming Zhang, Eric Zhu, and Themistocles Zikopoulos for helpful discussions. I am especially grateful to A. Khanikaev, S. Shirol, and E. Zhu for insightful comments on the experimental implementation, and to S. Sachdev for carefully reading the manuscript and providing valuable feedback.

I acknowledge the use of IBM Quantum services for this work. The views expressed are those of the author and do not reflect the official policy or position of IBM or the IBM Quantum team.

I also acknowledge the use of OpenAI's GPT-5.6 Sol through ChatGPT for assistance with manuscript editing, literature search, and checking analytical and numerical calculations. All outputs were reviewed and verified by the author, who takes full responsibility for the content of the manuscript.
\end{acknowledgments}

\bibliography{mybib}

\begin{center}
    \textbf{End Matter}
\end{center}

\begin{center}
    \textbf{Appendix A: Exact lattice solution and screening sum rules}
\end{center}
For Wilson polynomials with parameters $(a,b,c,d)=(1/2,1/2,3/2,3/2)$, we choose the normalization
\begin{equation}
    p_n(\lambda^2)=(-1)^n \sqrt{\frac{2n+3}{3}}\,\frac{ W_n(\lambda^2; \frac12,\frac12,\frac32,\frac32) }{n![(n+1)!]^2},
\end{equation}
with $p_0=1$. The normalized spectral density is
\begin{equation}
    \varrho(\lambda)=\frac{3}{2\pi}\left| \frac{\Gamma(1/2 + i\lambda)^2 \Gamma(3/2 + i\lambda)^2}{\Gamma(2i\lambda)} \right|^2,
\end{equation}
so that $\int_0^\infty d\lambda\, \varrho(\lambda) p_n(\lambda^2) p_m(\lambda^2) = \delta_{nm}$.

To derive the screening sum rule, factorize the lattice Hamiltonian as $H=BB^\dagger$, where the nonzero matrix elements of the matrix $B$ are
\begin{align}
    B_{nn} &  = \sqrt{\frac{(n+1)(n+2)(n+3)}{2(2n+3)}}, \\
    B_{n,n-1}&= \sqrt{\frac{n(n+1)(n+2)}{2(2n+3)}}.
\end{align}
Its inverse gives
\begin{equation}\label{eq:h-inverse}
    (H^{-1})_{mn}=(-1)^{m+n} \frac{\sqrt{(2m+3)(2n+3)}}{(N+1)(N+2)}.
\end{equation}
Here $N=\max(m,n)$. For generalized eigenstates of $W$ with $\braket{n}{w} = \sqrt{2n+3} P_n^{(0,2)}(2w-1)$ and $\mathbf1 = \int_0^1dw\,w^2\ket w \bra w$, Eq.~(\ref{eq:h-inverse}) yields 
\begin{equation}
    \bra w H^{-1} \ket v = -\frac{\log[1-\min(w,v)]}{w^2v^2}.
\end{equation}
The first inverse moment is therefore
\begin{align}
    \sum_j\Lambda_j^{-1} & =\frac{1}{4}\Tr(H^{-1}W^{2/d}) \nonumber \\ &= \frac{1}{4}\int_0^1dw\,w^{2/d-2}[-\log(1-w)] \nonumber
    \\ & =\frac{\psi(2/d)+\gamma_{\mathrm E}}{4(2/d-1)},
\end{align}
which gives Eq.~(\ref{eq:sum-rule}) and, for $d=4$, $\sum_j(2\pi T/M_j)^2=\log 2$. Higher inverse moments follow from the same kernel; an explicit second-moment result is given in the Supplemental Material.

\begin{center}
    \textbf{Appendix B: Finite-density continuum limit}
\end{center}
For a chain of $N$ sites at fixed filling $\nu=N_f/N$, define $u=n/N$. The large-$n$ coefficients give
\begin{equation}\label{eq:methods-semiclassical-energy}
    \mathcal E(u,q)=u^2\cos^2\frac{q}{2}.
\end{equation}
Taking $\mu_N=a^2N^2$, the local Fermi surface obeys
\begin{equation}\label{eq:methods-qF}
    \cos\frac{q_F(u)}{2}=\frac{a}u,\qquad u\ge a,
\end{equation}
while the region $u<a$ is completely filled. Integrating the local occupation gives
\begin{equation}\label{eq:methods-filling}
    \nu=\frac{2}{\pi}\left[\arcsin a+a\,\operatorname{arcosh}\frac{1}a\right],
\end{equation}
which fixes $a$ for a specified filling. Combining Eq.~(\ref{eq:methods-qF}) with the radial dictionary gives
\begin{equation}\label{eq:methods-rhoF}
    \rho_F(u)=2\operatorname{arcosh}\frac{u}a.
\end{equation}

Linearizing Eq.~(\ref{eq:methods-semiclassical-energy}) about the two local Fermi points gives
\begin{equation}\label{eq:methods-vF}
    v_F(u)=a\sqrt{u^2-a^2}.
\end{equation}
Since the microscopic one-particle energies scale as $N^2$, the finite-density low-energy limit is obtained from $H_{\mu_N}/N$. To leading order in gradients,
\begin{align}\label{eq:methods-dirac}
    H =\int du\,\bigg[ &-i\psi_R^\dagger \left(v_F\partial_u + \frac{1}{2} \partial_u v_F \right)\psi_R  \\ &+ i\psi_L^\dagger \left( v_F\partial_u+\frac12\partial_u v_F \right)\psi_L \bigg]. 
\end{align}
Introducing $d\xi/du=1/v_F(u)$ gives
\begin{equation}\label{eq:methods-xi}
    \xi(u)=\frac{1}a\operatorname{arcosh}\frac{u}a=\frac{\rho_F(u)}{2a}.
\end{equation}
After the field redefinition $\chi_{R,L}=\sqrt{v_F}\,\psi_{R,L}$, Eq.~(\ref{eq:methods-dirac}) becomes the homogeneous massless Dirac Hamiltonian
\begin{equation}
    H=\int d\xi\,(-i\chi_R^\dagger \partial_\xi\chi_R + i\chi_L^\dagger \partial_\xi \chi_L),
\end{equation}
which is a $c=1$ CFT. The point $u=a$ is the semiclassical turning point where the local linearization breaks down; away from it, the CFT description applies. Since $\ell=R\rho/d$, Eq.~(\ref{eq:methods-xi}) gives $\xi=d\ell/(2aR)$.

\clearpage
\onecolumngrid
\setcounter{section}{0}
\setcounter{subsection}{0}
\setcounter{equation}{0}
\setcounter{figure}{0}
\setcounter{table}{0}
\renewcommand{\thesection}{S\arabic{section}}
\renewcommand{\thesubsection}{\thesection.\arabic{subsection}}
\renewcommand{\theequation}{S\arabic{equation}}
\renewcommand{\theHequation}{S\arabic{equation}}
\renewcommand{\thefigure}{S\arabic{figure}}
\renewcommand{\theHfigure}{S\arabic{figure}}
\renewcommand{\thetable}{S\arabic{table}}

\begin{center}
{\large\bfseries Supplemental Material: Black hole radial geometry from a quantum spin chain}\\ 
Boris A. Khanikati\\
\small \textit{Department of Physics, Harvard University, Cambridge MA 02138, USA}
\end{center}

\section{Additional screening results}

\subsection{Lowest screening modes and BTZ check}
The generalized eigenvalue problem Eq.~(\ref{eq:screening-mass}) can be solved directly on a finite lattice and extrapolated with increasing chain length. In $d=4$, the lowest eigenvalue converges to $\Lambda_0\approx2.8969$, corresponding to
\begin{equation}
    \frac{M_0}{\pi T}=2\sqrt{\Lambda_0}\approx3.4041.
\end{equation}
This agrees with the previously obtained lowest screening mass in the massless-scalar channel of strongly coupled $\mathcal N=4$ supersymmetric Yang--Mills plasma \cite{bak2007screening}.

An exactly solvable check is $d=2$, corresponding to the nonrotating Ba\~nados--Teitelboim--Zanelli (BTZ) black hole. For a scalar operator of dimension $\Delta$, the static poles of the finite-temperature two-point function occur at \cite{birmingham2002cft}
\begin{equation}
    M_j=4\pi T\left(j+\frac\Delta2\right),\qquad j=0,1,2,\ldots.
\end{equation}
For a massless bulk scalar, $\Delta=2$, so
\begin{equation}
    \frac{M_j}{\pi T}=4(j+1),\qquad \Lambda_j=4(j+1)^2,
\end{equation}
which is reproduced by Eq.~(\ref{eq:screening-mass}). Hence
\begin{equation}
    \sum_{j=0}^\infty\Lambda_j^{-m}=\frac{\zeta(2m)}{4^m},
\end{equation}
and the first inverse moment is $\pi^2/24$, consistent with the $d\to2$ limit of Eq.~(\ref{eq:sum-rule}).

\subsection{Higher inverse moments}
For $d=4$, the kernel in Appendix A also gives
\begin{align}
    \sum_j\Lambda_j^{-2} &= \frac1{16} \Tr(H^{-1}W^{1/2}H^{-1}W^{1/2}) \nonumber \\ & =\frac14\int_0^1dw\,\log^2(1-w)\left(\frac1{w^2} -\frac1{w^{3/2}}\right),
\end{align}
and therefore
\begin{equation}
    \sum_{j=0}^\infty\left(\frac{2\pi T}{M_j}\right)^4 = 2\log^2{2} -\frac{\pi^2}{12}.
\end{equation}

\subsection{Asymptotic screening spectrum}
The large-$n$ structure of the chain gives $\epsilon_n\simeq n^2/2$, $t_n\simeq n^2/4$, so locally $H(n,q)\simeq n^2\cos^2(q/2)$ and $W^{1/2}(n,q)\simeq\cos(q/2)$. The generalized eigenvalue equation then gives
\begin{equation}
    \Lambda\simeq4n^2\cos^{2-4/d}\frac q2.
\end{equation}
For fixed $\Lambda$, the semiclassical region is bounded by
\begin{equation}
    n<n_0(q),\qquad n_0(q)=\frac{\sqrt\Lambda}{2\cos^{1-2/d}(q/2)}.
\end{equation}
Phase-space counting gives
\begin{equation}
    \mathcal N(\Lambda) \simeq \frac{\Gamma(1/d)}{2\sqrt\pi\,\Gamma(1/2+1/d)} \sqrt\Lambda,
\end{equation}
so the large-$j$ screening masses have asymptotically constant spacing,
\begin{equation}
    \frac{M_{j+1}-M_j}{\pi T}\to4\sqrt\pi\frac{\Gamma(1/2+1/d)}{\Gamma(1/d)}.
\end{equation}
This agrees with the known WKB asymptotics of the continuum radial problem \cite{Minahan1999}.

\section{Quantum implementation and reconstruction}

\begin{figure*}[t!]
\centering
\includegraphics[width=\linewidth]{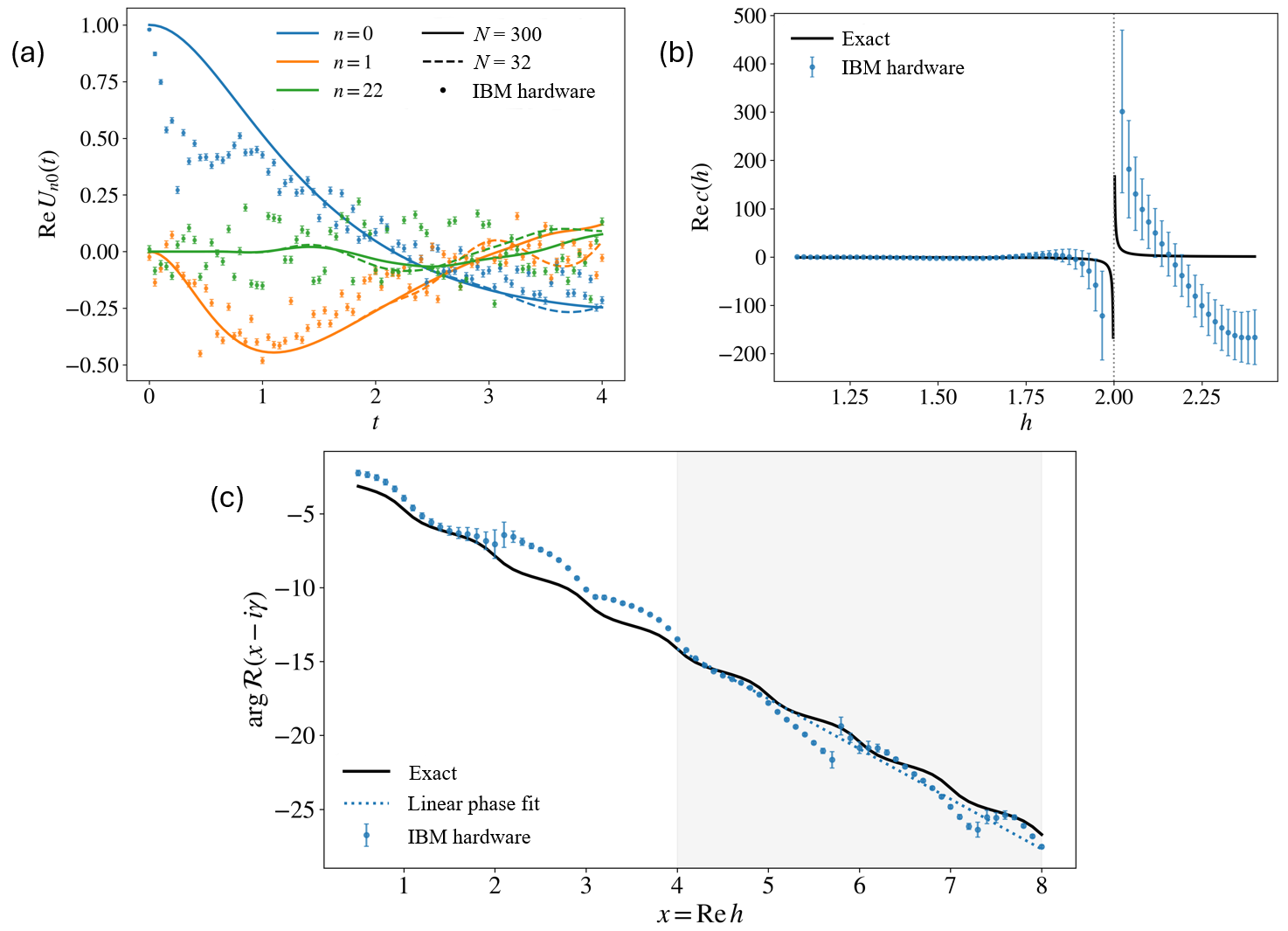}
\caption{Measurements and reconstruction of observables. (a) Real part of the source-to-lattice propagator $U_{n0}(\tau)$ for representative lattice sites. Quantum-hardware measurements (points) are compared with the numerically exact $N=32$ truncated-chain evolution (dashed lines) and an $N=300$ approximation to the semi-infinite-chain result (solid lines).
(b) Thermal one-point coefficient $c(h)$ reconstructed from the quantum-hardware data, compared with the exact black brane result (solid black line). The vertical dotted line marks the first pole at $h=2$. (c) Unwrapped phase of the response $\mathcal R(h)$ along $h=x-i\gamma$, with $\gamma=0.2$. The solid black line is the exact black brane result, while the dotted line is a weighted linear fit to the experimental phase over $4\le x\le8$. The fit yields $\tau_s/R=0.847\pm0.017$, compared with the exact value $\pi/4\simeq0.785$. Error bars denote $1\sigma$ finite-shot statistical uncertainties.}
\label{fig:Fig_2}
\end{figure*}

\subsection{Spin-chain implementation}
Up to an additive constant, Eq.~(\ref{eq:main-hamiltonian}) can be written as
\begin{equation}
    H=-\frac12\sum_n\epsilon_nZ_n+\frac12\sum_nt_{n+1}(X_nX_{n+1}+Y_nY_{n+1}).
\end{equation}
The computational vacuum $\ket{0\cdots0}$ represents $\ket{\Omega}$, and $\ket{n}$ is the state with a single $1$ at site $n$. Excitation number is conserved, allowing the one-particle dynamics to be compiled with number-conserving orbital rotations. The experiment truncates the chain to $N=32$ sites. For each simulator time $\tau$, the single-particle unitary $U_1(\tau)=e^{-iH_N\tau}$ is calculated classically and implemented as an \texttt{ffsim} spinless Jordan--Wigner orbital rotation, with the \texttt{ffsim} pre-initialization transpiler pass applied before backend transpilation \cite{ffsim2026}.

A Hadamard gate on qubit $0$ prepares $(\ket{\Omega}+\ket{0})/\sqrt2$. Because the vacuum is invariant under the number-conserving evolution, it provides a phase reference. After evolution,
\begin{equation}
    \ket{\Psi(\tau)} = \frac{1}{\sqrt2} \left[ \ket{\Omega} + \sum_nU_{n0}(\tau)\ket{n} \right],
\end{equation}
so that
\begin{equation}
    U_{n0}(\tau)=\langle X_n(\tau)\rangle+i\langle Y_n(\tau)\rangle.
\end{equation}
All qubits are therefore measured globally in the $X$ and $Y$ bases.

The source propagator is sampled at
\begin{equation}
\tau_j=j\Delta\tau,\qquad \Delta\tau=0.05,\qquad 0\le\tau_j\le4.0,
\end{equation}
for $N_t=81$ simulator times, with 4096 shots per measurement circuit. Circuits are transpiled at Qiskit optimization level 3 with transpiler seed 42, and the \texttt{XpXm} dynamical-decoupling sequence is applied \cite{qiskit2024}. The 162 response circuits were submitted in nine IBM Runtime jobs; the maximum transpiled depth was 344 and the maximum number of two-qubit gates was 992.

Figure~\ref{fig:Fig_2}a compares the measured propagator with the numerically exact $N=32$ truncated-chain evolution and an $N=300$ approximation to the semi-infinite chain. The hardware data reproduce the qualitative propagation across the chain, while deviations increase at larger $n$. The displayed error bars include finite-shot statistics only; systematic hardware errors are not included.

\subsection{Static-response reconstruction}
The frequency-domain response is reconstructed from the measured real-time propagator using the finite-time transform
\begin{equation}
G^{(\tau_{\max})}_{n0}(\zeta)\simeq i\sum_ja_je^{i\zeta\tau_j}U_{n0}(\tau_j),
\end{equation}
where $a_j$ are trapezoidal quadrature weights. For
\begin{equation}
\zeta(h)=\frac14-\left(h-\frac12\right)^2,
\end{equation}
the reconstructed radial response is
\begin{equation}
\phi_n(h)=-\frac{B_d}{\sqrt3}G^{(\tau_{\max})}_{n0}[\zeta(h)].
\end{equation}
The normalizable coefficient is extracted by fitting to
\begin{equation}\label{eq:methods-bn}
b_n(h)=\sqrt{2n+3}\,\frac{\Gamma(h-1)\Gamma(h+1)}{\Gamma(h-n-1)\Gamma(h+n+2)}
\end{equation}
over the primary window $8\le n\le22$. We extract $c(h)$ using a weighted complex least-squares fit,
\begin{equation}
\widehat c(h)=\frac{\sum_n\sigma_{\phi,n}^{-2}b_n^*(h)\phi_n(h)}{\sum_n\sigma_{\phi,n}^{-2}|b_n(h)|^2},\qquad
\sigma_c=\left[\sum_n\sigma_{\phi,n}^{-2}|b_n(h)|^2\right]^{-1/2}.
\end{equation}
The fitting window is chosen to probe the asymptotic regime while avoiding the largest sites, where finite-chain and experimental deviations are most pronounced; stability is checked by varying the window.

Finite-shot uncertainties are propagated through the same transform,
\begin{equation}
\sigma_{G,n}^2(\zeta)=\sum_j|a_je^{i\zeta\tau_j}|^2(\sigma_{X,jn}^2+\sigma_{Y,jn}^2),\qquad
\sigma_{\phi,n}=\frac{|B_d|}{\sqrt3}\sigma_{G,n}.
\end{equation}
These uncertainties do not include systematic hardware errors. The same finite-time transform and fitting procedure are applied to the numerically exact $N=32$ truncated-chain propagator to diagnose finite-chain and finite-time effects. The $N=300$ reference in Fig.~\ref{fig:Fig_2}a approximates the semi-infinite-chain evolution over the experimental time window. Figure~\ref{fig:Fig_2}b compares the experimentally reconstructed $c(h)$ with the exact black brane result, Eq.~(\ref{eq:one-point}); agreement deteriorates beyond the first pole, where the reconstruction is especially sensitive to finite-chain, finite-time, and experimental errors.

\subsection{Interior-phase extraction}
For the analytic continuation, $c(h)$ is reconstructed along
\begin{equation}
    h=x-i\gamma,\qquad \gamma=0.2,
\end{equation}
for $0.5\le x\le8$. We form
\begin{equation}
    \mathcal R(h) = c(h)\frac{\Gamma(2h)}{\Gamma(h)^2h(1-h)}
\end{equation}
and unwrap its phase as a function of $x$. The solid black curve in Fig.~\ref{fig:Fig_2}c shows the exact black brane result, $\mathcal R(h)=-\pi B_d/\sin(\pi h)$.

The phase uncertainty is estimated by
\begin{equation}
    \sigma_{\arg\mathcal R} \simeq \frac{\sigma_{\mathcal R}}{|\mathcal R|}, \qquad \sigma_{\mathcal R} = \left|\frac{\Gamma(2h)}{\Gamma(h)^2h(1-h)}\right| \sigma_c.
\end{equation}
A weighted linear fit to the experimental phase over $4\leq x \leq 8$, shown as the dotted line in Fig.~\ref{fig:Fig_2}c, uses
\begin{equation}
    \arg\mathcal R(x-i\gamma)=-\alpha x+\varphi_0,
\end{equation}
with $\tau_s/R=\alpha/d$. The fit gives $\tau_s/R=0.847\pm0.017$, compared with the exact black brane value $\pi/4\simeq0.785$. The quoted uncertainty reflects finite-shot statistics only. The deviation from the exact value is larger than this statistical uncertainty and therefore indicates systematic contributions not included in the error bar. Finite-size, finite-time, and reconstruction effects are assessed using numerically exact finite-chain calculations and by varying the fitting window.

\end{document}